\documentclass[useAMS,a4paper,fleqn,usenatbib]{mnras}

\usepackage[T1]{fontenc}
\usepackage{ae,aecompl}

\usepackage{graphicx}	
\usepackage{times}
\usepackage{amsmath}	
\usepackage{amssymb}	
\usepackage{tabularx}
\usepackage{txfonts}
\usepackage{wasysym}
\usepackage{ragged2e}
\usepackage{natbib}
\usepackage[english]{babel}
\usepackage{float}
\usepackage{color}

\usepackage{aas}

\title[Variable stars in the OC M\,67]{Variable stars in one open cluster
  within the \textit{Kepler/K2-Campaign-5} field: M\,67 (NGC\,2682)\thanks{ Based on
    observations collected with the Schmidt\,67/92 Telescope at the
    Osservatorio Astronomico di Asiago, which is part of the
    Osservatorio Astronomico di Padova, Istituto Nazionale di
    Astrofisica.  }\thanks{Light curves of variable stars available at
    http://groups.dfa.unipd.it/\-ESPG/\-aphn.html\,. } }

\author[Nardiello et al.]{D.\ Nardiello$^{1,2}$\thanks{E-mail: domenico.nardiello@unipd.it}, 
M.\ Libralato$^{1,2}$, 
L.\ R.\ Bedin$^{1}$, 
G.\ Piotto$^{1,2}$,
P.\ Ochner$^{1}$, 
A.\ Cunial$^{1,2}$,
\newauthor
L.\ Borsato$^{1,2}$, 
V.\ Granata$^{1,2}$\\ 
$^{1}$Istituto Nazionale di Astrofisica - Osservatorio Astronomico di Padova, Vicolo dell'Osservatorio 5, Padova, IT-35122 \\
$^{2}$Dipartimento di Fisica e Astronomia ``Galileo Galilei'', Universit\`a di Padova, Vicolo dell'Osservatorio 3, Padova IT-35122 \\
}

\date{Accepted 2015 October 17. Received 2015 October 3; in original form 2015 October 3.}

\pubyear{2015}

\begin{document}
\label{firstpage}
\pagerange{\pageref{firstpage}--\pageref{lastpage}}
\maketitle

\begin{abstract}
In this paper we continue the release of high-level data products from
the multiyear photometric survey collected at the 67/92~cm Schmidt
Telescope in Asiago.
The primary goal of the survey is to discover and to characterise
variable objects and exoplanetary transits in four fields containing
five nearby open clusters spanning a broad range of ages.

This second paper releases a photometric catalogue, in five
photometric bands, of the Solar-age, Solar-metallicity open cluster
M\,67 (NGC\,2682).  Proper motions are derived comparing the positions
observed in 2013 at the Asiago's Schmidt Telescope with those
extracted from WFI@2.2m MPG/ESO images in 2000.
We also analyse the variable sources within M\,67.  We detected 68
variables, 43 of which are new detection.  Variable periods and
proper-motion memberships of a large majority of sources in our
catalogue are improved with respect to previous releases. The entire
catalogue will be available in electronic format.

Besides the general interest on an improved catalogue, this work will
be particularly useful because of: (1) the imminent release of {\it
  Kepler/K2 Campaign-5} data of this clusters, for which our catalogue
will provide an excellent, high spatial resolution input list, and (2)
characterisation of the M\,67 stars which are targets of intense
HARPS and HARPS-N radial-velocity surveys for planet search.
\end{abstract}



\section{Introduction}

In \citet[hereafter Paper\,I]{2015MNRAS.447.3536N} we presented our
multi-year, multi-wavelength photometric survey programme: ``The
\mbox{Asiago} Pathfinder for HARPS-N'' (hereafter, APHN; PI:~Bedin)
aimed at characterising variable stars and transiting-exoplanet
candidates in five open clusters (OCs).
Originally, APHN was intended as a preparatory survey for the on-going
searches of planets in OCs with the High Accuracy Radial velocity
Planet Searcher for the Northern hemisphere (HARPS-N) mounted at the
Telescopio Nazionale Galileo (TNG).
The APHN survey has also recently acquired further interest, as four
out of the five monitored OCs were chosen as targets for the {\it
  Kepler} extended mission {\it
  K2}\footnote{http://keplerscience.arc.nasa.gov/K2/}
(\citealt{2014PASP..126..398H}).

In Paper\,I we analysed the two OCs M\,35 and NGC\,2158, for which we
released atlases and stacked images.
In this second paper we present the third of our target OCs: M\,67
(NGC\,2682).
The OC M\,67 has been the subject of many investigations. Among them:
the search and study of variable stars
(\citealt{1991AJ....101..541G,2002A&A...382..899S,2002A&A...382..888V,2003AJ....126.2954S,2003AJ....125.2173S,2006MNRAS.373.1141S,2007MNRAS.377..584S,2007MNRAS.378.1371B,2008MNRAS.391..343P,2009A&A...503..165Y}),
the determinations of proper motions and memberships
(\citealt{1977A&AS...27...89S,1989AJ.....98..227G,1993A&AS..100..243Z,2008A&A...484..609Y,2015AJ....150...97G}),
radial velocities
(\citealt{1986AJ.....92.1100M,1986AJ.....92.1364M,1990AJ....100.1859M,1992PASP..104.1268M,1994AJ....108.1828M,2011A&A...526A.127P,2015AJ....150...97G}),
the estimates of cluster parameters (age $\sim$4\,Gyr, distance
$<$1\,kpc, metallicity [Fe/H]$\sim$0, $E(B-V)\sim 0.05$; \citealt{
  1984AJ.....89..487J, 1986AJ.....92.1364M, 1987AJ.....93..634N,
  1992AJ....103..151D, 1993AJ....106..181M, 1994A&AS..103..375C,
  1996AJ....112..628F, 2003MNRAS.345.1015G, 2004PASP..116..997V,
  2004MNRAS.347..101S, 2006A&A...450..557R, 2007A&A...470..585B,
  2007AJ....133..370T, 2008A&A...489..677P, 2009ApJ...698.1872S,
  2010A&A...511A..56P, 2011AJ....142...59J}),
the search for exoplanets candidates
(\citealt{2012A&A...545A.139P,2014A&A...561L...9B}),
and the dynamical studies
(\citealt{2010A&A...513A..51B,2012AJ....143...73P}).

Our investigation is focused on finding new variable stars and
extracting a complete astro-photometric catalogue (with improved
proper motions and membership probabilities) in a region of $\sim
0.6$\,degree$^2$, centred M\,67.

The paper is organised as follows. In Section~\ref{obser} we describe
our data set, the data reduction, the extraction and the detrending of
the light curves. In Section~\ref{varfind} we show the tools used for
finding variable stars. Section~\ref{varcmd} is dedicated to variable
star detection, proper motions and membership probabilities
computation, and colour-magnitude diagrams (CMDs) presentation.  In
Section~\ref{em} we describe the released electronic
material. Finally, in Section~\ref{summ} there is a summary of our
work.

\section{Observations, data reduction, light curves extraction and detrending}
\label{obser}

\begin{figure*}
\includegraphics[width=\hsize]{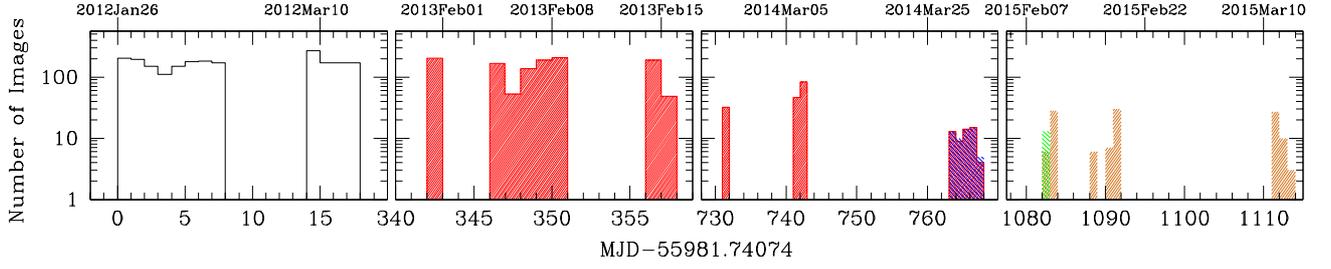}
\caption{ Histogram of the number of images per night collected during
  the four campaigns of our programme.  The white histogram refers to
  observations in white light; the red, blue, green, and mango
  histograms refer to observations in the $R$, $B$, $V$, and $I$
  filters, respectively.  }
\label{obs}
\end{figure*}

All images of the OC M\,67 [($\alpha, \delta $)=($08^{\rm h}51^{\rm
    m}18^{\rm s},+11^{\circ}48^{'}00^{''}$)] were collected with the
Asiago 67/92\,cm Schmidt Telescope located on Mount Ekar (longitude
$11.\!^{\circ}5710$~E, latitude $45.\!^{\circ}8430$~N, altitude
1370\,m), that belongs to the Astronomical Observatory of Padova
(OAPD), which is part of the Istituto Nazionale di Astrofisica
(INAF). At the focus of the Schmidt telescope there is a SBIG
STL-11000M camera, equipped with a Kodak KAI-11000M detector
($4050\times2672$ pixel, field of view: $58\times38$\,arcmin$^2$,
pixel scale: 862.5\,mas\,pixel$^{-1}$).  The characteristics of the
telescope and the CCD are described in details in Paper\,I.

The OC M\,67 is one of the four fields observed under the long-term
observing programme APHN (PI:~Bedin), aimed to characterise variable
stars and transiting-exoplanet candidates in five OCs.

As in the case of M35 and NGC\,2158 (Paper\,I), in the first observing
season (2012) M\,67 data were collected in white light (hereafter
indicated with filter $N$, where $N$ stands for 'None'), with exposure
time of 120\,s, and 60\,s (during the almost-full moon nights); during
the second (2013) and the third season (2014) we collected
180\,s+15\,s $R$-filter and 180\,s $B$-filter images. Finally, during
the fourth season (2015), the observations were carried out in
$I$-band (240\,s+15\,s) and $V$-band (240\,s). In table \ref{tab:log}
we give a log of the observations, while in Fig.~\ref{obs} we show the
histograms of the number of images taken during all the observing
seasons.

All images are stored in the INAF national archive in
Trieste\footnote{http://ia2.oats.inaf.it/archives/asiago}.

\begin{table}
\caption{Log of observations.}
\medskip
\label{tab:log}
\begin{tabular}{l c c c c}
\hline
{\bf Filter} & {\bf \# Images} & {\bf Exp. Time}    & {\bf FWHM}  & {\bf Median FWHM}  \\
             &           &  (s)               &  (arcsec)   &  (arcsec)          \\
\hline 
\hline
$B$          &      57   &    $180$          & 1.20--1.86  & 1.47               \\
$V$          &      13   &    $240$          & 1.59--2.31  & 1.90               \\
$R$          &     588   &    $15$           & 1.33--7.66  & 3.17               \\
             &     822   &    $180$          &             &                    \\
$I$          &      39   &    $15$           & 1.44--6.16  & 2.98               \\
             &      78   &    $240$          &             &                    \\
$N$          &     232   &    $ 60$          & 1.70--7.38  & 3.41               \\
             &    1878   &    $120$          &             &                    \\
\hline
\end{tabular}

\justify
\end{table}

For the data reduction, we used the software described in detail in
Paper\,I.  For each image we made a grid of $9 \times 5$
spatially-varying, empirical point spread functions (PSFs), one for
each of the 45 regions in which we divided the field of view (FOV);
the software is similar to that developed by
\citet{2006A&A...454.1029A} for the Wide Field Imager (WFI) mounted at
the ESO / Max-Planck-Gesellschaft (MPG) 2.2m telescope.

For any location on the detector, the best PSF model is obtained by a
bilinear interpolation of the four closest PSFs and it is used to
measure the position and the flux of the stars in each image.

We transformed stellar positions in all the images into the reference
frame of the best image\footnote{The ``best image'' is characterised
  by the minimum of the product between airmass and seeing.} in filter
$N$ (ID \texttt{SC02906}). For each filter, we obtained the
photometric zero-points between the single images and the best image
in that filter (ID: \texttt{SC02906} for $N$-filter, \texttt{SC37438}
for $R$-filter, \texttt{SC37437} for $B$-filter, \texttt{SC46181} for
$V$-filter, \texttt{SC46722} for $I$-filter).

\begin{figure*}
\begin{minipage}[c]{\textwidth}
\includegraphics[width=\hsize]{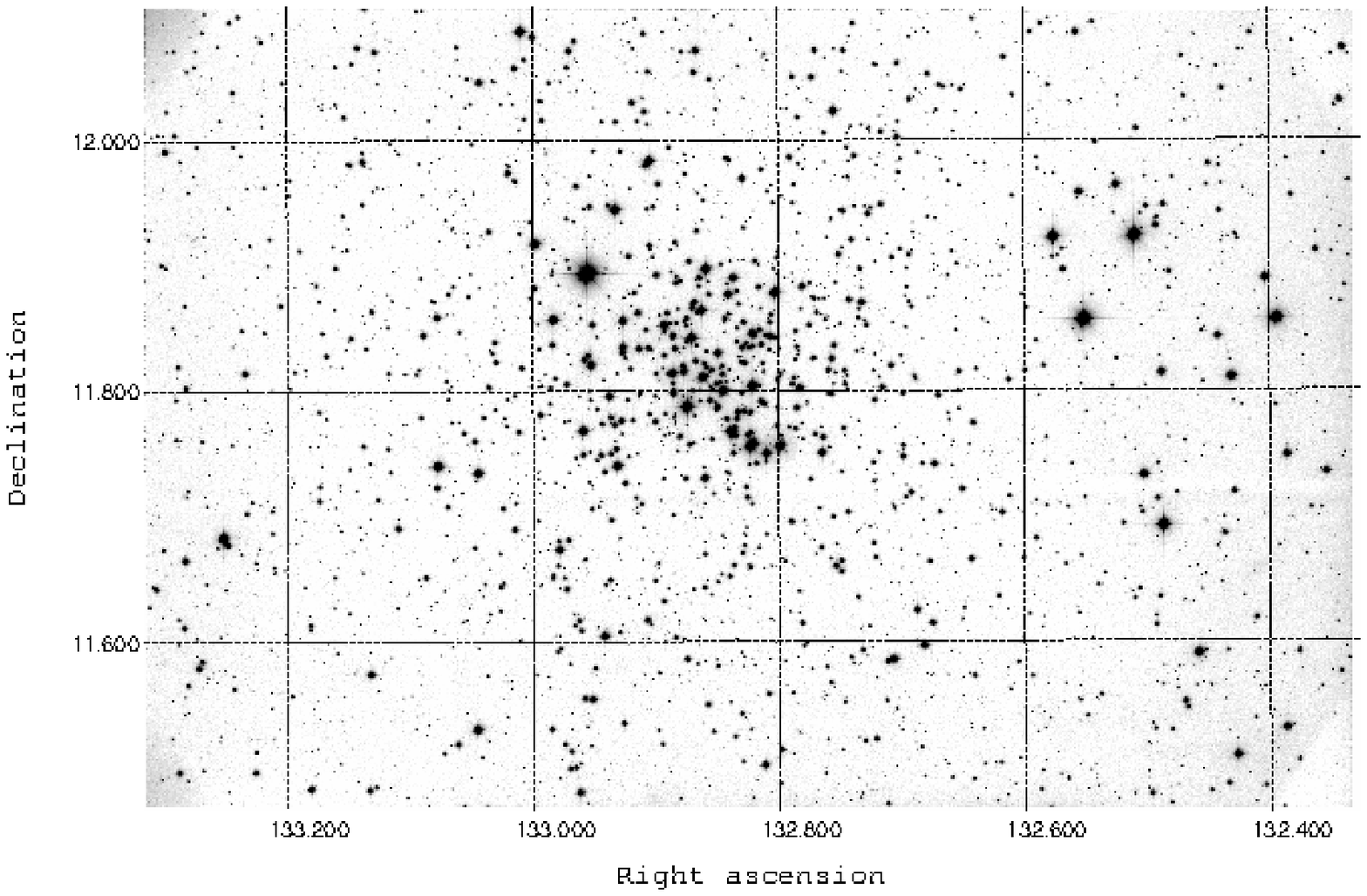}
\end{minipage}
\begin{minipage}[c]{\textwidth}
\vspace{3mm}
\centering
\hspace{10mm}
\includegraphics[width=0.28\hsize]{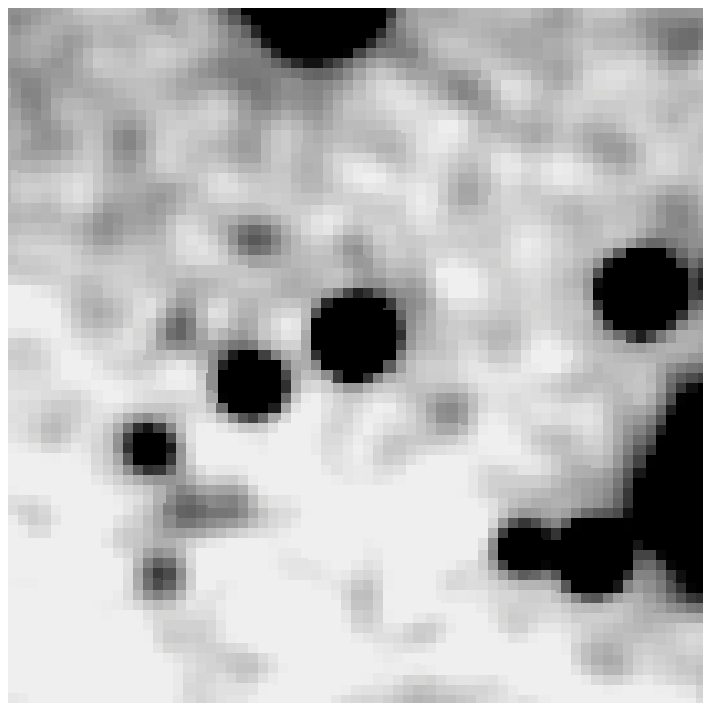} 
\includegraphics[width=0.28\hsize]{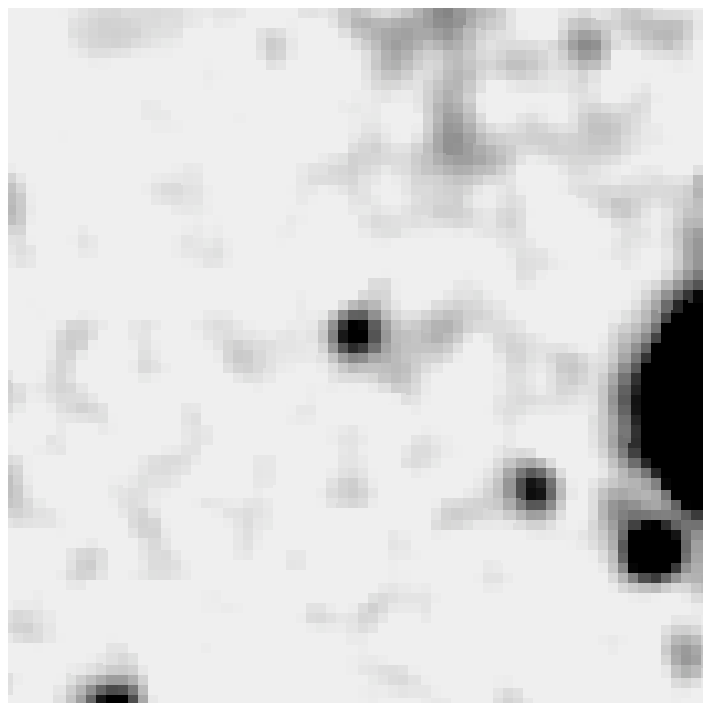} 
\includegraphics[width=0.28\hsize]{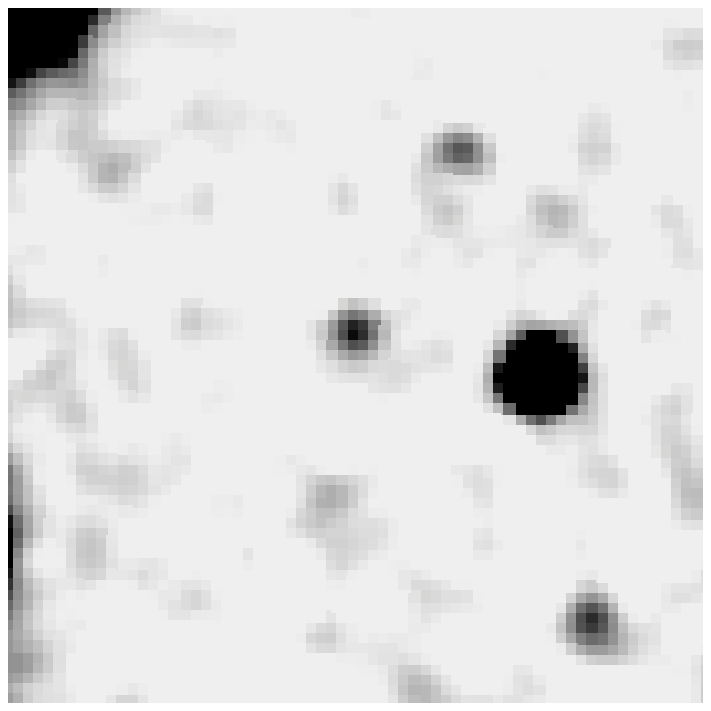} 
\end{minipage}
\caption{{\it Top:} $N$-filter stacked image of M\,67 obtained with
  our data. {\it Bottom:} zoom-in of the $N$-filter stacked image
  centred on the three white dwarfs shown in Fig.~2 of
  \citet{2010A&A...513A..50B}.  }
\label{fov}
\end{figure*}

We created a stacked, high-signal-to-noise ratio (SNR) image (`stack')
of the field for each filter. Using the $N$-filter stack,
characterised by the highest SNR (Fig.~\ref{fov}), and the software
used for single images, we derived an improved star list.  We purged
this star list from false detection using the \texttt{qfit} parameter
(a diagnostic related to the quality of the PSF fit,
\citealt{2008AJ....135.2114A}) and the procedure described in
\citet{2014A&A...563A..80L}.  The purged star list contains 6905
sources, and we used it as the master star list (MSL) for the
extraction of the light curves (LCs).  The MSL contains stars with $V
\lesssim 25$; faint enough to reach the magnitude of the faintest
white dwarfs (WDs) detected along the cooling sequence of M\,67 by
\citet{2010A&A...513A..50B}.  The three bottom panels of
Fig.~\ref{fov} are centred on the three WDs shown in Fig.~2 of
\citet{2010A&A...513A..50B}. The bottom-right panel of Fig.~\ref{fov}
shows the faintest WD ($V\sim 24$) identified by the same authors.
We extracted the $BVRI$-photometry of the 6905 sources in our
catalogue by using the $BVRI$-stacked images. We calibrated the
magnitudes by matching our catalogue to the Stetson Standard star
catalogue (\citealt{2000PASP..112..925S}). We derived calibration
equations by means of least squares fitting of straight lines using
magnitudes and colours.

\begin{figure*}
\includegraphics[width=\hsize]{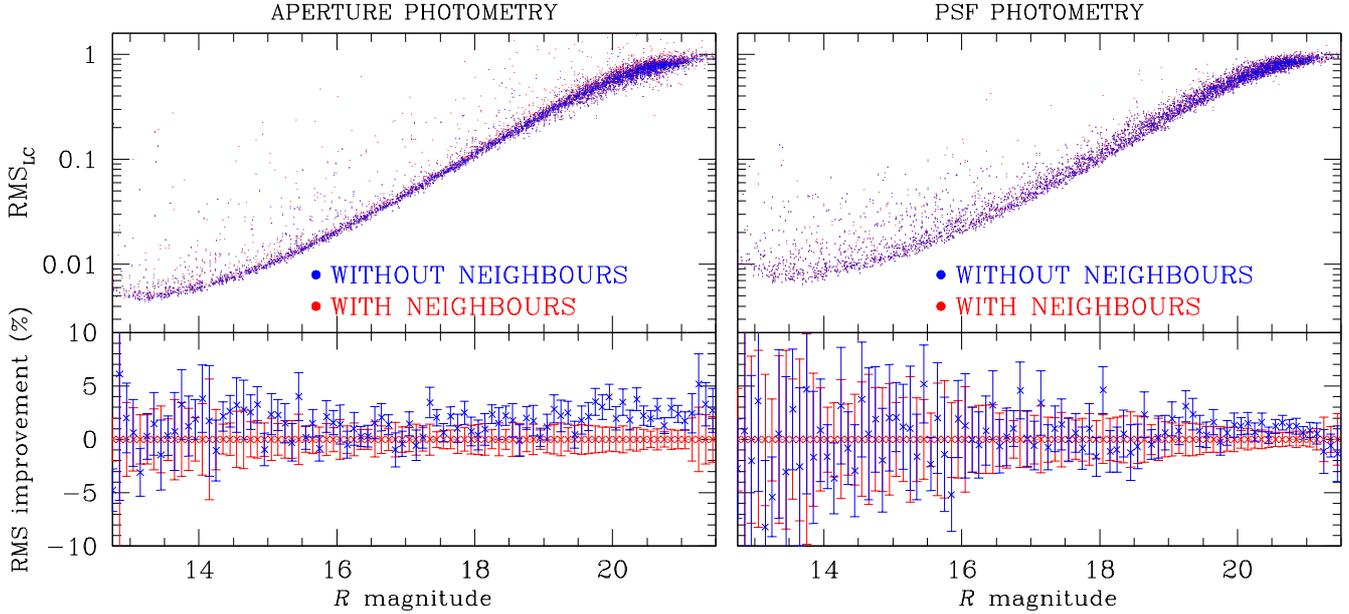}
\caption{
Photometric rms, obtained from original 180s $R$ images (red) and from
images after neighbour-subtraction (blue), for both aperture
photometry (left) and PSF photometry (right). The lower panels show
the percentage improvement of the neighbour-subtraction algorithm on
the rms, averaged over 0.1-mag bins.
}
\label{rms1}
\end{figure*}

\begin{figure}
\includegraphics[width=\hsize]{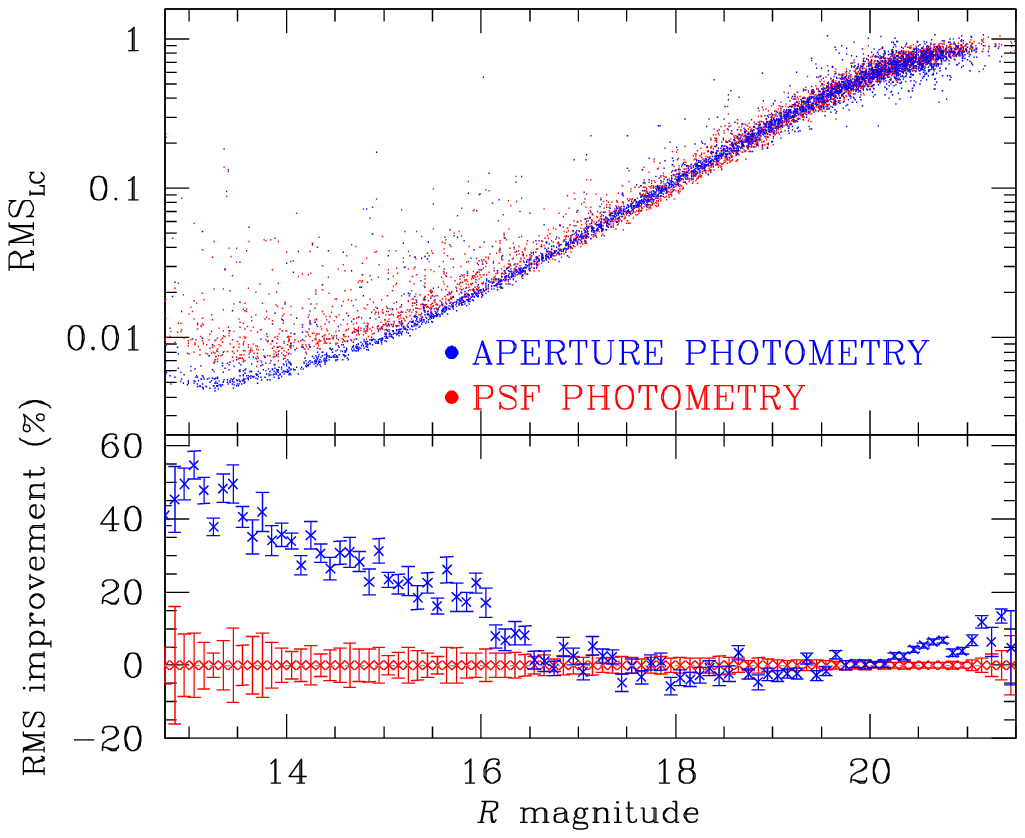}
\caption{
As in Fig.~\ref{rms1}, but comparing aperture and PSF photometries on
neighbour-subtracted images.
}
\label{rms2}
\end{figure}

For the LCs extraction, we used the software developed and described
in details in Paper\,I. Briefly, by using (i) the MSL catalogue, (ii)
the PSFs, and (iii) the six-parameters linear transformations between
the MSL and the single-image catalogues, for each target star within
the MSL, the software extracts LCs in two parallel versions: a first
one from the original images, and a second one from images where the
neighbours close to the target star were PSF-fitted and subtracted. In
both versions, our tool extracts the flux of the target star using PSF
fitting and aperture photometry. As in Paper\,I, for aperture
photometry we adopted a dynamical aperture that depends on the Full
Width Half Maximum (FWHM), with radius $r$=$1.0\times{\rm
  FWHM}$. Therefore, for each star in the MSL, four light curves are
generated: PSF with/without neighbour-subtracted and aperture
with/without neighbour-subtracted.

In order to remove residual systematic errors from the LCs, we
followed the procedure used in Paper\,I and described in detail in
\citet{2014MNRAS.442.2381N}.  For each target star in each image, our
algorithm computes local, weighted photometric zero-points using
selected reference stars (generally the stars with the best rms at a
given magnitude). The weights are a function of the relative on-sky
position and of the magnitude difference between the target and the
reference star.  The software also computes a global zero-point
correction, which ---on average--- provides worse residuals than the
local zero-point correction, as also found in Paper\,I.

In Fig.~\ref{rms1} we show the photometric rms of the LCs derived from
all the photometric methods (aperture with/without neighbours and
PSF-fitting with/without neighbours).
Even if the field of M\,67 is relatively loose (if compared to the
field analysed in Paper\,I) the photometry on images after neighbours
subtraction is better than the photometry on original images, as the
rms-scatter is lower and the rms-improvement is -- on average -- $\sim
3$\% in the case of aperture photometry, and $\sim 1$\% in the case of
PSF-fitting photometry.
Images were slightly de-focused to avoid saturation of main sequence
(MS) turn off stars, therefore it was difficult to perfectly model the
PSFs. This is also the reason why the rms PSF-fitting photometry is
overall worse than the one associated to aperture photometry, as shown
in Fig.~\ref{rms2}.

\section{Variable finding}
\label{varfind}

\begin{figure*}
\includegraphics[width=\hsize]{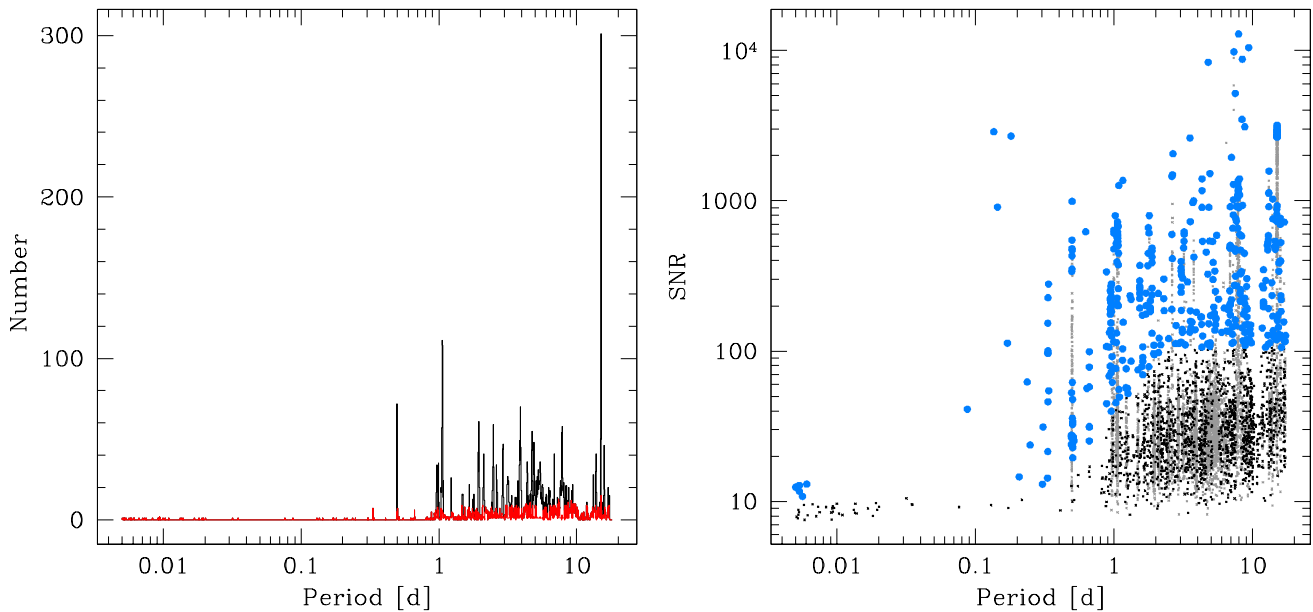}
\caption{
Procedure followed for the extraction of candidate variable stars.
Left panel: distribution of the periods obtained from all the light
curves before (black) and after (red) spikes suppression. Right
panel:the AoV SNR as a function of the period of the light curves: in
grey all the stars, in black the stars after spikes suppression, and
in azure the suspected variables.
}
\label{find}
\end{figure*}

In order to detect candidate variables in our field, we used three
different algorithms: the Generalized Lomb-Scargle (GLS) periodogram
(\citealt{2009A&A...496..577Z}), suitable for sinusoidal signals; the
Analysis of Variance (AoV) periodogram
(\citealt{1989MNRAS.241..153S}), useful to detect all variable types;
the Box-fitting Least-Squares (BLS) periodogram
(\citealt{2002A&A...391..369K}), effective for searching box-like dips
in an otherwise flat or nearly flat LC, such as eclipsing binaries
(EBs) and planetary transits. All the algorithms are part of the
publicly available code \texttt{VARTOOLS}
v.1.32\footnote{http://www.astro.princeton.edu/$\sim$jhartman/vartools.html}
(\citealt{2008ApJ...675.1254H}). We used the output parameters
associated to each algorithm for excluding the sources in our
catalogue that have low probabilities to be variable in our data.  In
order to isolate the candidate variable stars, we used the procedure
described in Paper\,I and summarised in Fig.~\ref{find}.

From the histograms of the detected periods for all the LCs, we
removed the spikes (saving the stars with high SNR, left panel of
Fig.~\ref{find}). The spikes are associated to spurious periods due to
systematic effects, such as instrumental and atmospheric artefacts. In
a second step, we selected by hand the candidate variables in the SNR
versus period diagram (right panel of Fig.~\ref{find}), and we
visually inspected each of them.

We performed this procedure on neighbour-subtracted LCs, both for
aperture and PSF photometries and for $N$ and $R$ LCs, identifying 68
real variable sources.

As in Paper\,I, we refined the periods using the following procedure:
for each variable star, we normalised the $R$ and $N$ LC to zero,
subtracting the 5$\sigma$-clipped median magnitude. Then, we merged
the $R$ and $N$ normalised LCs, obtaining a LC with a temporal
baseline of 764 days.  We used again the \texttt{VARTOOLS} algorithms
LS, AoV, and BLS to improve the period of the variable star. This
procedure is useful only to improve the periods; it is not possible to
extract any other scientific information from this normalised LC.

\section{Variable stars and colour-magnitude diagrams}
\label{varcmd}

In our catalogue there are 6905 sources that cover a field-of-view of
$\sim 0.612$ degree$^2$ centred on M\,67. In this catalogue we find 68
variables stars. All the variable stars are listed in
Table~\ref{tab:var}; for each variable we provide the identification
number (ID), the position, the period (if it is not irregular), and,
when available, the magnitudes in $NBVRIJ_{\rm 2MASS}H_{\rm
  2MASS}K_{\rm 2MASS}$ bands, the membership probabilities as obtained
in Sect.~\ref{memb}, and the radial velocities from
\citet{2015AJ....150...97G}.

Of these 68 variable stars, 25 variable stars have already been
classified in other photometric works
(\citealt{1991AJ....101..541G,2002A&A...382..899S,2002A&A...382..888V,2003AJ....126.2954S,2003AJ....125.2173S,2006MNRAS.373.1141S,2007MNRAS.377..584S,2007MNRAS.378.1371B,2008MNRAS.391..343P,2009A&A...503..165Y})
and/or in the General Catalogue of Variable Stars (GCVS,
\citealt{2013A&AT...28...49S}). Other variable stars listed in
literature catalogues, but not found in our variable catalogue, are
bright objects extremely saturated in our data (even in short
exposures) or just outside the Schmidt FOV.

\begin{table*}
\caption{First 10 lines of the catalogue of variable stars.}
\medskip
\label{tab:var}
\footnotesize
\begin{tabular}{l c c c c c c c c c c c c c }
\hline
ID & $\alpha$ & $\delta$ & P     & $N$ & $B$ & $V$ & $R$ & $I$ & $J_{\rm 2MASS}$ &  $H_{\rm 2MASS}$ & $K_{\rm 2MASS}$ & $P_\mu$      & RV$^{a}$ \\ 
   & (degree) & (degree) & (day) &     &     &     &     &     &               &               &               &  (\%)   & (km\,s$^{-1}$) \\
(1) & (2) & (3) & (4) & (5) & (6) & (7) & (8) & (9) & (10) & (11) & (12) & (13) & (14) \\
\hline
  10  & 132.48057 & 11.470550 & 6.61403975 & -13.4283 & 15.6986 & 15.8312 & 14.8366 & 14.7724 & 13.665 & 13.276 & 13.262 & 90.5418 & 30.29\\
  37  & 132.72530 & 11.480997 & 1.82830531 & -10.526 & 20.3047 & 18.5991 & 17.7994 & 15.9833 & 14.631 & 13.999 & 13.724 & 2.3146 & -1000.00\\
  101 & 133.03346 & 11.503185 & 0.61441096 & -9.9831 & 19.6606 & 18.757 & 18.0689 & 17.6468 & 16.401 & 15.96 & 15.59 & 14.8792 & -1000.00  \\
  142 & 132.54111 & 11.514107 & 0.33958032 & -9.3075 & 19.7992 & 19.3732 & 18.9384 & 18.9665 & -99.999 & -99.999 & -99.999 & 44.9397 &-1000.00 \\
  144 & 132.68428 & 11.515616 & 8.07609064 & -13.5201 & 15.783 & 14.9577 & 14.7485 & 14.6447 & 13.572 & 13.167 & 13.079 & 0.347 & 55.34\\
  193 & 133.08555 & 11.534070 & 5.01456608 & -11.7733 & 17.9946 & 16.8831 & 16.339 & 15.7722 & 14.592 & 13.965 & 13.791 & 97.3684 & -1000.00 \\
  207 & 133.09478 & 11.541483 & 10.8060368 & -9.0553 & -99.999 & 18.7736 & 18.9221 & 18.3115 & 16.559 & 15.826 & 15.243 & 49.6983 & -1000.00\\
  211 & 132.54888 & 11.540289 & 19.1807153 & -14.0378 & 15.3067 & 14.4544 & 14.146 & 13.971 & 12.883 & 12.444 & 12.341 & 0.0 & 35.69\\
  236 & 132.46906 & 11.551604 & 2.86317050 & -16.0982 & 13.3535 & 12.4298 & 12.0711 & 11.9546 & 10.725 & 10.269 & 10.18 & 0.0 & -15.61\\
  239 & 132.55463 & 11.552562 & 6.75840156 & -13.0745 & 16.1264 & 15.4163 & 15.1784 & 15.0959 & 14.076 & 13.766 & 13.733 & 0.0 & 13.22\\
\hline\end{tabular}

\justify
\footnotesize
{\it Notes.}\,\,\,$^a$\citet{2015AJ....150...97G}.
\end{table*}

\subsection{Proper motions and membership probabilities}
\label{memb}

We used stellar proper motions to separate cluster members and field
stars. The approach adopted to compute stellar positional displacement
is the same as in many other works, e.g.,
\citet{2003AJ....126..247B,2006A&A...454.1029A,2008A&A...484..609Y,2010A&A...513A..50B,2015A&A...573A..70N},
and \citet{2015MNRAS.450.1664L}.  Using six-parameter local
transformations and a sample of likely cluster members (for example MS
stars), we computed the displacement between the stellar positions in
two different epochs, after been transformed into a common reference
system. As first epoch, we used M\,67 $V$-filter observations
collected with the WFI mounted at the ESO/MPG 2.2\,m telescope, on
February 16th, 2000 ($t_{\rm I}$=2000.1). As second epoch we used the
best 100 Schmidt $R$-images collected during the 2013 observational
run ($t_{\rm II}\simeq$2013.1). The time baseline for the proper
motion measurements is $\sim$13.0\,yr.

\begin{figure*}
\centering
\includegraphics[width=\hsize]{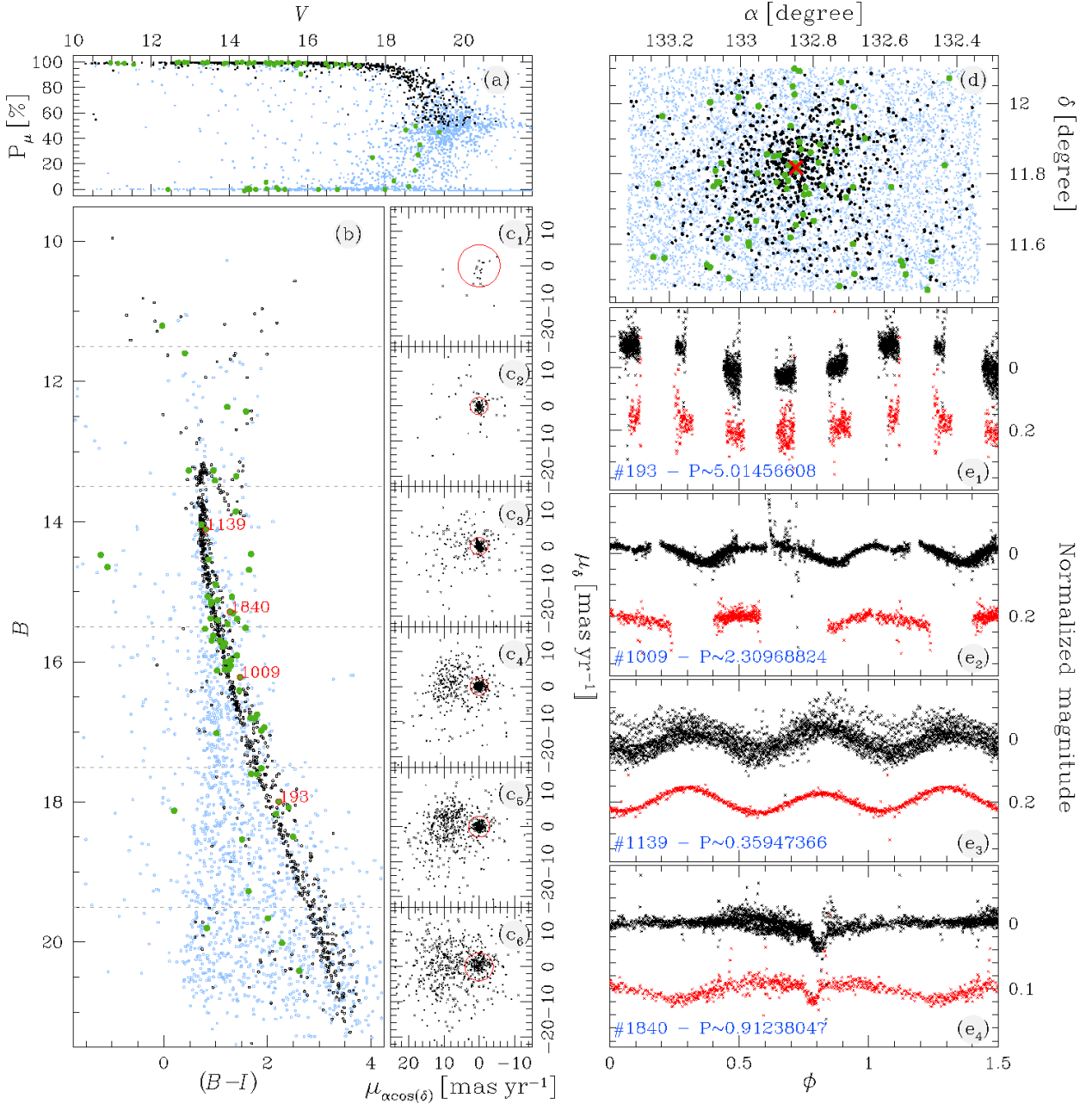}
\caption{ Panel (a): membership probabilities $P_\mu$ as a function of
  the $V$ magnitude; panel (b): CMD $B$ versus ($B-I$) of M\,67 stars
  in the MSL; panels (c): vector-point diagrams for the same stars as in
  panel (b) in the corresponding magnitude intervals. The cluster star proper motion
  distribution is centred around (0,0) ; panel (d): positions
  ($\alpha$,$\delta$) for all the stars in the Schmidt MSL. The red
  cross indicates the cluster centre (\citealt{2008A&A...484..609Y});
  panels (e): four examples of median-magnitude normalised LCs in
  filter $N$ (black) and $R$ (red). The four stars have
  $P_\mu>97$\%.  In panels (a), (b), and (d), we plot in black the
  likely cluster members, i.e. the stars inside the red circles in
  panels (c), in azure the likely field stars, and in green the
  variable stars found in this work. In panel (b) we highlight the
  variable stars shown in panels (e).
}
\label{cmds}
\end{figure*}

Since we used likely cluster members to compute the coefficients of
the six-parameter linear transformations, the stellar displacements
are computed \textit{relative} to the cluster mean motion. Therefore,
by construction, the cluster distribution in the vector-point diagram
(VPD) is centred around (0,0), while field stars, that have a
different motion relative to that of the cluster, lie in a different
region of the VPD (see Fig.~\ref{cmds}).

The membership probabilities (MPs) of the stars in the M\,67 FOV have
already been calculated using both astrometry
(e.g. \citealt{1977A&AS...27...89S,1989AJ.....98..227G,1993A&AS..100..243Z,2008A&A...484..609Y})
and radial velocities (e.g
\citealt{2008A&A...484..609Y,2011A&A...526A.127P,2015AJ....150...97G}).
Our final catalogue supplements the other works; we extracted the MPs
in an homogeneous way for stars with $V\lesssim 19.5$ in a region of
$58\times38$\,arcmin$^2$.

There are different methods to derive stellar MPs. We chose that
described by \citet{1998A&AS..133..387B}. Briefly, we derived the
frequency function for both cluster ($\Phi_c$) and field stars
($\Phi_f$). We assumed that cluster distribution is centred at
($\mu_{\alpha\cos\delta,c}$,$\mu_{\delta,c}$) $=$ (0.00,0.00) mas
yr$^{-1}$ with an intrinsic dispersion\footnote{The intrinsic
  dispersion is dominated by the positional errors and do not reflect
  the true intrinsic cluster dispersion.} of
($\sigma_{\mu_{\alpha\cos\delta,c}}$,$\sigma_{\mu_{\delta,c}}$) $=$
(0.65,0.71) mas yr$^{-1}$. For field stars, we have
($\mu_{\alpha\cos\delta,f}$,$\mu_{\delta,f}$) $=$ (9.34,$-$1.10) mas
yr$^{-1}$ and a dispersion of
($\sigma_{\mu_{\alpha\cos\delta,f}}$,$\sigma_{\mu_{\delta,f}}$) $=$
(4.19,6.19) mas yr$^{-1}$. We ignored the spatial distribution of the
two components and assumed that there is not a correlation between
them (correlation coefficient $\gamma$ is set to 0). We excluded from
our calculation poorly-measured-proper-motion stars. The membership
probability is then computed as:

\begin{displaymath}
  P_\mu = \frac{\Phi_c}{\Phi_c + \Phi_f} \,\,.
\end{displaymath}

Panel (a) of Fig.~\ref{cmds} shows the available MPs $P_\mu$ as a
function of the magnitude $V$. The figure shows that we measured
reliable MPs for stars with $V\lesssim19.5$.

\subsection{Colour-magnitude diagrams and light curves}

In panel (b) of Fig.~\ref{cmds} we show the $B$ versus $(B-I)$ CMD of
M\,67. In panels (a) and (b), we plot in black the likely cluster
member stars, selected according to their proper motions (i.e. the
  stars inside the red circles in the VPD of panels c), in azure the stars
rejected and those are likely field stars, and with green dots the
variable stars found in this work. In panel (d), using the same
colour-code, we plot the positions of the stars in our MSL field of view.

In panels (e) there are four examples of variable stars with
$P_\mu>97$\%. The LCs are normalised to the median magnitude. We plot
in black the LC in $N$-filter and in red that in $R$-filter.

Panel (e$_4$) shows an eclipsing binary (ID\,\#1840) that has
never been photometrically observed, but that is in the sample of
\citet{2015AJ....150...97G}. From spectroscopic observations, they
found that this is a double lined (SB2) eclipsing binary and also a
X-ray source (X35 in the {\it ROSAT} catalogue,
\citealt{1998A&A...339..431B}). In our data, we probably observe only
the primary eclipse, that has a box-like shape with a depth of
0.02\,mag.

\section{Electronic material}
\label{em}

The catalogue of all the sources in our MSL is electronically
available. The catalogue contains the following information: Cols~(1)
and (2) are the J2000.0 equatorial coordinates in decimal degrees;
Cols~(3) and (4) the positions in pixel on the $N$-filter stack;
Cols~(5)-(12) are the instrumental $N$, and the calibrated $BVRIJ_{\rm
  2MASS}H_{\rm 2MASS}K_{\rm 2MASS}$ magnitudes (when the magnitude is
not available, it is flagged with -99.999); Col.~(13) is the
identification number (ID) of the star; Cols.~(14) and (15) give the
relative proper motions in mas\,yr$^{-1}$ along ($\alpha \cos \delta,
\delta$) direction (when it is not available, it is flagged with
-999.9999); Col.~(16) gives the membership probabilities $P_\mu$ (when it
is not available, it is flagged with $-1$).

Along with the MSL, we release the catalogue of all variable stars
found in this work. An example of the catalogue is
Table~\ref{tab:var}: Col.~(1) is the identification number (ID) of the
variable star in our MSL; Cols~(2) and (3) are the J2000.0 equatorial
coordinates in decimal degree; Col.~(4) contains the periods in day
(when the star is irregular, the period is -99); Cols (5)-(12) are the
instrumental $N$, and the calibrated $BVRIJ_{\rm 2MASS}H_{\rm
  2MASS}K_{\rm 2MASS}$ magnitudes (when the magnitude is not
available, it is flagged with -99.999); Col.~(13) gives the membership
probabilities $P_\mu$ (when it is not available, it is flagged with
$-1$); Col.~(14) are the radial velocities measured by
\citet{2015AJ....150...97G} (when the radial velocity measurement is
not available, it is denoted as -1000). For each variable star, we
also release the $BRIN$ LCs.

Finally, we make the astrometrized stacks in $BVRIN$ filters
electronically available.

\section{Summary}
\label{summ}

In this work we present the photometric results for the third target (M\,67)
of the photometric survey of OC stars conducted with the 67/92\,cm
Schmidt telescope at Mount Ekar, Asiago. We analysed a total of 3707
images in $B$, $V$, $R$, $I$ and white light (no filter), collected
over 3.1~years. We used the algorithms described in
\citet{2015MNRAS.447.3536N} to obtain a complete list of stars (6905
sources with magnitude $V \lesssim 25$) and to extract, detrend, and
analyse the corresponding LCs.  We
identify 68 variable stars (43 of which are new).  Combining Schmidt
best data with WFI@2.2m MPG/ESO images (collected on 2000), we derived
the relative proper motions and the membership probabilities for
a great number of stars in our MSL.  We release two electronic
catalogues: the catalogue of variable stars, (containing
coordinates, periods $B$, $V$, $R$, $I$, $N$, 2MASS magnitudes,
membership probabilities, radial velocities) and the catalogue of all
the sources in the Schmidt FOV (containing positions, $B$, $V$,
$R$, $I$, $N$, 2MASS magnitudes, proper motions, membership
probabilities). The electronic material includes the $B$, $V$, $R$,
$I$, and white light stacked images and light curves of the identified
variable stars.

The OC M\,67 is in the field of the {\it K2 Mission-Campaign-5}. The
released catalogue of M\,67 sources will be an excellent
input-list for the extraction of LCs from {\it K2} images. In this
sense, our survey is preparatory to the analysis of {\it K2} data, and
complements (and extends in time) the light curves of the stars covered
by {\it K2}.

\section*{Acknowledgements}
We warmly thank the referee, Dr. R. Gilliland, for the prompt and
careful reading of our manuscript. \\
DN, ML, LRB, GP, AC, LB, and VG acknowledge PRIN-INAF 2012 partial
funding under the project entitled: ``The M\,4 Core Project with
Hubble Space Telescope''. DN and GP also acknowledge partial support
by the Universit\`a degli Studi di Padova Progetto di Ateneo CPDA141214
``Towards understanding complex star formation in Galactic globular
clusters''.

\bibliographystyle{mn2e} \bibliography{biblio}

\begin{thebibliography}{59}
\expandafter\ifx\csname natexlab\endcsname\relax\def\natexlab#1{#1}\fi

\bibitem[{{Anderson} {et~al}\mbox{.}(2006){Anderson}, {Bedin}, {Piotto},
  {Yadav}, \& {Bellini}}]{2006A&A...454.1029A}
{Anderson} J., {Bedin} L.~R., {Piotto} G., {Yadav} R.~S., {Bellini} A., 2006,
  \aap, 454, 1029

\bibitem[{{Anderson} {et~al}\mbox{.}(2008){Anderson}, {King}, {Richer},
  {Fahlman}, {Hansen}, {Hurley}, {Kalirai}, {Rich}, \&
  {Stetson}}]{2008AJ....135.2114A}
{Anderson} J. {et~al.}, 2008, \aj, 135, 2114

\bibitem[{{Balaguer-N{\'u}{\~n}ez}, {Galad{\'{\i}}-Enr{\'{\i}}quez} \&
  {Jordi}(2007){Balaguer-N{\'u}{\~n}ez}, {Galad{\'{\i}}-Enr{\'{\i}}quez}, \&
  {Jordi}}]{2007A&A...470..585B}
{Balaguer-N{\'u}{\~n}ez} L., {Galad{\'{\i}}-Enr{\'{\i}}quez} D., {Jordi} C.,
  2007, \aap, 470, 585

\bibitem[{{Balaguer-N{\'u}nez}, {Tian} \& {Zhao}(1998){Balaguer-N{\'u}nez},
  {Tian}, \& {Zhao}}]{1998A&AS..133..387B}
{Balaguer-N{\'u}nez} L., {Tian} K.~P., {Zhao} J.~L., 1998, \aaps, 133, 387

\bibitem[{{Bedin} {et~al}\mbox{.}(2003){Bedin}, {Piotto}, {King}, \&
  {Anderson}}]{2003AJ....126..247B}
{Bedin} L.~R., {Piotto} G., {King} I.~R., {Anderson} J., 2003, \aj, 126, 247

\bibitem[{{Bellini} {et~al}\mbox{.}(2010{\natexlab{a}}){Bellini}, {Bedin},
  {Pichardo}, {Moreno}, {Allen}, {Piotto}, \& {Anderson}}]{2010A&A...513A..51B}
{Bellini} A., {Bedin} L.~R., {Pichardo} B., {Moreno} E., {Allen} C., {Piotto}
  G., {Anderson} J., 2010{\natexlab{a}}, \aap, 513, A51

\bibitem[{{Bellini} {et~al}\mbox{.}(2010{\natexlab{b}}){Bellini}, {Bedin},
  {Piotto}, {Salaris}, {Anderson}, {Brocato}, {Ragazzoni}, {Ortolani},
  {Bonanos}, {Platais}, {Gilliland}, {Raimondo}, {Bragaglia}, {Tosi},
  {Gallozzi}, {Testa}, {Kochanek}, {Giallongo}, {Baruffolo}, {Farinato},
  {Diolaiti}, {Speziali}, {Carraro}, \& {Yadav}}]{2010A&A...513A..50B}
{Bellini} A. {et~al.}, 2010{\natexlab{b}}, \aap, 513, A50

\bibitem[{{Belloni}, {Verbunt} \& {Mathieu}(1998){Belloni}, {Verbunt}, \&
  {Mathieu}}]{1998A&A...339..431B}
{Belloni} T., {Verbunt} F., {Mathieu} R.~D., 1998, \aap, 339, 431

\bibitem[{{Brucalassi} {et~al}\mbox{.}(2014){Brucalassi}, {Pasquini}, {Saglia},
  {Ruiz}, {Bonifacio}, {Bedin}, {Biazzo}, {Melo}, {Lovis}, \&
  {Randich}}]{2014A&A...561L...9B}
{Brucalassi} A. {et~al.}, 2014, \aap, 561, L9

\bibitem[{{Bruntt} {et~al}\mbox{.}(2007){Bruntt}, {Stello}, {Su{\'a}rez},
  {Arentoft}, {Bedding}, {Bouzid}, {Csubry}, {Dall}, {Dind}, {Frandsen},
  {Gilliland}, {Jacob}, {Jensen}, {Kang}, {Kim}, {Kiss}, {Kjeldsen}, {Koo},
  {Lee}, {Lee}, {Nuspl}, {Sterken}, \& {Szab{\'o}}}]{2007MNRAS.378.1371B}
{Bruntt} H. {et~al.}, 2007, \mnras, 378, 1371

\bibitem[{{Carraro} {et~al}\mbox{.}(1994){Carraro}, {Chiosi}, {Bressan}, \&
  {Bertelli}}]{1994A&AS..103..375C}
{Carraro} G., {Chiosi} C., {Bressan} A., {Bertelli} G., 1994, \aaps, 103, 375

\bibitem[{{Demarque}, {Green} \& {Guenther}(1992){Demarque}, {Green}, \&
  {Guenther}}]{1992AJ....103..151D}
{Demarque} P., {Green} E.~M., {Guenther} D.~B., 1992, \aj, 103, 151

\bibitem[{{Fan} {et~al}\mbox{.}(1996){Fan}, {Burstein}, {Chen}, {Zhu}, {Jiang},
  {Wu}, {Yan}, {Zheng}, {Zhou}, {Fang}, {Chen}, {Deng}, {Chu}, {Hester},
  {Windhorst}, {Li}, {Lu}, {Sun}, {Chen}, {Tsay}, {Chiueh}, {Chou}, {Ko},
  {Lin}, {Guo}, \& {Byun}}]{1996AJ....112..628F}
{Fan} X. {et~al.}, 1996, \aj, 112, 628

\bibitem[{{Geller}, {Latham} \& {Mathieu}(2015){Geller}, {Latham}, \&
  {Mathieu}}]{2015AJ....150...97G}
{Geller} A.~M., {Latham} D.~W., {Mathieu} R.~D., 2015, \aj, 150, 97

\bibitem[{{Gilliland} {et~al}\mbox{.}(1991){Gilliland}, {Brown}, {Duncan},
  {Suntzeff}, {Lockwood}, {Thompson}, {Schild}, {Jeffrey}, \&
  {Penprase}}]{1991AJ....101..541G}
{Gilliland} R.~L. {et~al.}, 1991, \aj, 101, 541

\bibitem[{{Girard} {et~al}\mbox{.}(1989){Girard}, {Grundy}, {Lopez}, \& {van
  Altena}}]{1989AJ.....98..227G}
{Girard} T.~M., {Grundy} W.~M., {Lopez} C.~E., {van Altena} W.~F., 1989, \aj,
  98, 227

\bibitem[{{Grocholski} \& {Sarajedini}(2003)}]{2003MNRAS.345.1015G}
{Grocholski} A.~J., {Sarajedini} A., 2003, \mnras, 345, 1015

\bibitem[{{Hartman} {et~al}\mbox{.}(2008){Hartman}, {Gaudi}, {Holman},
  {McLeod}, {Stanek}, {Barranco}, {Pinsonneault}, \&
  {Kalirai}}]{2008ApJ...675.1254H}
{Hartman} J.~D., {Gaudi} B.~S., {Holman} M.~J., {McLeod} B.~A., {Stanek} K.~Z.,
  {Barranco} J.~A., {Pinsonneault} M.~H., {Kalirai} J.~S., 2008, \apj, 675,
  1254

\bibitem[{{Howell} {et~al}\mbox{.}(2014){Howell}, {Sobeck}, {Haas}, {Still},
  {Barclay}, {Mullally}, {Troeltzsch}, {Aigrain}, {Bryson}, {Caldwell},
  {Chaplin}, {Cochran}, {Huber}, {Marcy}, {Miglio}, {Najita}, {Smith},
  {Twicken}, \& {Fortney}}]{2014PASP..126..398H}
{Howell} S.~B. {et~al.}, 2014, \pasp, 126, 398

\bibitem[{{Jacobson}, {Pilachowski} \& {Friel}(2011){Jacobson}, {Pilachowski},
  \& {Friel}}]{2011AJ....142...59J}
{Jacobson} H.~R., {Pilachowski} C.~A., {Friel} E.~D., 2011, \aj, 142, 59

\bibitem[{{Janes} \& {Smith}(1984)}]{1984AJ.....89..487J}
{Janes} K.~A., {Smith} G.~H., 1984, \aj, 89, 487

\bibitem[{{Kov{\'a}cs}, {Zucker} \& {Mazeh}(2002){Kov{\'a}cs}, {Zucker}, \&
  {Mazeh}}]{2002A&A...391..369K}
{Kov{\'a}cs} G., {Zucker} S., {Mazeh} T., 2002, \aap, 391, 369

\bibitem[{{Libralato} {et~al}\mbox{.}(2015){Libralato}, {Bellini}, {Bedin},
  {Anderson}, {Piotto}, {Nascimbeni}, {Platais}, {Minniti}, \&
  {Zoccali}}]{2015MNRAS.450.1664L}
{Libralato} M. {et~al.}, 2015, \mnras, 450, 1664

\bibitem[{{Libralato} {et~al}\mbox{.}(2014){Libralato}, {Bellini}, {Bedin},
  {Piotto}, {Platais}, {Kissler-Patig}, \& {Milone}}]{2014A&A...563A..80L}
{Libralato} M., {Bellini} A., {Bedin} L.~R., {Piotto} G., {Platais} I.,
  {Kissler-Patig} M., {Milone} A.~P., 2014, \aap, 563, A80

\bibitem[{{Mathieu} \& {Latham}(1986)}]{1986AJ.....92.1364M}
{Mathieu} R.~D., {Latham} D.~W., 1986, \aj, 92, 1364

\bibitem[{{Mathieu}, {Latham} \& {Griffin}(1990){Mathieu}, {Latham}, \&
  {Griffin}}]{1990AJ....100.1859M}
{Mathieu} R.~D., {Latham} D.~W., {Griffin} R.~F., 1990, \aj, 100, 1859

\bibitem[{{Mathieu} {et~al}\mbox{.}(1986){Mathieu}, {Latham}, {Griffin}, \&
  {Gunn}}]{1986AJ.....92.1100M}
{Mathieu} R.~D., {Latham} D.~W., {Griffin} R.~F., {Gunn} J.~E., 1986, \aj, 92,
  1100

\bibitem[{{Milone}(1992)}]{1992PASP..104.1268M}
{Milone} A.~A.~E., 1992, \pasp, 104, 1268

\bibitem[{{Milone} \& {Latham}(1994)}]{1994AJ....108.1828M}
{Milone} A.~A.~E., {Latham} D.~W., 1994, \aj, 108, 1828

\bibitem[{{Montgomery}, {Marschall} \& {Janes}(1993){Montgomery}, {Marschall},
  \& {Janes}}]{1993AJ....106..181M}
{Montgomery} K.~A., {Marschall} L.~A., {Janes} K.~A., 1993, \aj, 106, 181

\bibitem[{{Nardiello} {et~al}\mbox{.}(2015{\natexlab{a}}){Nardiello}, {Bedin},
  {Nascimbeni}, {Libralato}, {Cunial}, {Piotto}, {Bellini}, {Borsato},
  {Brogaard}, {Granata}, {Malavolta}, {Marino}, {Milone}, {Ochner}, {Ortolani},
  {Tomasella}, {Clemens}, \& {Salaris}}]{2015MNRAS.447.3536N}
{Nardiello} D. {et~al.}, 2015{\natexlab{a}}, \mnras, 447, 3536

\bibitem[{{Nardiello} {et~al}\mbox{.}(2015{\natexlab{b}}){Nardiello}, {Milone},
  {Piotto}, {Marino}, {Bellini}, \& {Cassisi}}]{2015A&A...573A..70N}
{Nardiello} D., {Milone} A.~P., {Piotto} G., {Marino} A.~F., {Bellini} A.,
  {Cassisi} S., 2015{\natexlab{b}}, \aap, 573, A70

\bibitem[{{Nascimbeni} {et~al}\mbox{.}(2014){Nascimbeni}, {Bedin}, {Heggie},
  {van den Berg}, {Giersz}, {Piotto}, {Brogaard}, {Bellini}, {Milone}, {Rich},
  {Pooley}, {Anderson}, {Ubeda}, {Ortolani}, {Malavolta}, {Cunial}, \&
  {Pietrinferni}}]{2014MNRAS.442.2381N}
{Nascimbeni} V. {et~al.}, 2014, \mnras, 442, 2381

\bibitem[{{Nissen}, {Twarog} \& {Crawford}(1987){Nissen}, {Twarog}, \&
  {Crawford}}]{1987AJ.....93..634N}
{Nissen} P.~E., {Twarog} B.~A., {Crawford} D.~L., 1987, \aj, 93, 634

\bibitem[{{Pancino} {et~al}\mbox{.}(2010){Pancino}, {Carrera}, {Rossetti}, \&
  {Gallart}}]{2010A&A...511A..56P}
{Pancino} E., {Carrera} R., {Rossetti} E., {Gallart} C., 2010, \aap, 511, A56

\bibitem[{{Pasquini} {et~al}\mbox{.}(2008){Pasquini}, {Biazzo}, {Bonifacio},
  {Randich}, \& {Bedin}}]{2008A&A...489..677P}
{Pasquini} L., {Biazzo} K., {Bonifacio} P., {Randich} S., {Bedin} L.~R., 2008,
  \aap, 489, 677

\bibitem[{{Pasquini} {et~al}\mbox{.}(2012){Pasquini}, {Brucalassi}, {Ruiz},
  {Bonifacio}, {Lovis}, {Saglia}, {Melo}, {Biazzo}, {Randich}, \&
  {Bedin}}]{2012A&A...545A.139P}
{Pasquini} L. {et~al.}, 2012, \aap, 545, A139

\bibitem[{{Pasquini} {et~al}\mbox{.}(2011){Pasquini}, {Melo}, {Chavero},
  {Dravins}, {Ludwig}, {Bonifacio}, \& {de La Reza}}]{2011A&A...526A.127P}
{Pasquini} L., {Melo} C., {Chavero} C., {Dravins} D., {Ludwig} H.-G.,
  {Bonifacio} P., {de La Reza} R., 2011, \aap, 526, A127

\bibitem[{{Pichardo} {et~al}\mbox{.}(2012){Pichardo}, {Moreno}, {Allen},
  {Bedin}, {Bellini}, \& {Pasquini}}]{2012AJ....143...73P}
{Pichardo} B., {Moreno} E., {Allen} C., {Bedin} L.~R., {Bellini} A., {Pasquini}
  L., 2012, \aj, 143, 73

\bibitem[{{Pribulla} {et~al}\mbox{.}(2008){Pribulla}, {Rucinski}, {Matthews},
  {Kallinger}, {Kuschnig}, {Rowe}, {Guenther}, {Moffat}, {Sasselov}, {Walker},
  \& {Weiss}}]{2008MNRAS.391..343P}
{Pribulla} T. {et~al.}, 2008, \mnras, 391, 343

\bibitem[{{Randich} {et~al}\mbox{.}(2006){Randich}, {Sestito}, {Primas},
  {Pallavicini}, \& {Pasquini}}]{2006A&A...450..557R}
{Randich} S., {Sestito} P., {Primas} F., {Pallavicini} R., {Pasquini} L., 2006,
  \aap, 450, 557

\bibitem[{{Samus} \& {Antipin}(2013)}]{2013A&AT...28...49S}
{Samus} N.~N., {Antipin} S.~V., 2013, Astronomical and Astrophysical
  Transactions, 28, 49

\bibitem[{{Sanders}(1977)}]{1977A&AS...27...89S}
{Sanders} W.~L., 1977, \aaps, 27, 89

\bibitem[{{Sandquist}(2004)}]{2004MNRAS.347..101S}
{Sandquist} E.~L., 2004, \mnras, 347, 101

\bibitem[{{Sandquist} \& {Shetrone}(2003{\natexlab{a}})}]{2003AJ....126.2954S}
{Sandquist} E.~L., {Shetrone} M.~D., 2003{\natexlab{a}}, \aj, 126, 2954

\bibitem[{{Sandquist} \& {Shetrone}(2003{\natexlab{b}})}]{2003AJ....125.2173S}
{Sandquist} E.~L., {Shetrone} M.~D., 2003{\natexlab{b}}, \aj, 125, 2173

\bibitem[{{Sarajedini}, {Dotter} \& {Kirkpatrick}(2009){Sarajedini}, {Dotter},
  \& {Kirkpatrick}}]{2009ApJ...698.1872S}
{Sarajedini} A., {Dotter} A., {Kirkpatrick} A., 2009, \apj, 698, 1872

\bibitem[{{Schwarzenberg-Czerny}(1989)}]{1989MNRAS.241..153S}
{Schwarzenberg-Czerny} A., 1989, \mnras, 241, 153

\bibitem[{{Stassun} {et~al}\mbox{.}(2002){Stassun}, {van den Berg}, {Mathieu},
  \& {Verbunt}}]{2002A&A...382..899S}
{Stassun} K.~G., {van den Berg} M., {Mathieu} R.~D., {Verbunt} F., 2002, \aap,
  382, 899

\bibitem[{{Stello} {et~al}\mbox{.}(2006){Stello}, {Arentoft}, {Bedding},
  {Bouzid}, {Bruntt}, {Csubry}, {Dall}, {Dind}, {Frandsen}, {Gilliland},
  {Jacob}, {Jensen}, {Kang}, {Kim}, {Kiss}, {Kjeldsen}, {Koo}, {Lee}, {Lee},
  {Nuspl}, {Sterken}, \& {Szab{\'o}}}]{2006MNRAS.373.1141S}
{Stello} D. {et~al.}, 2006, \mnras, 373, 1141

\bibitem[{{Stello} {et~al}\mbox{.}(2007){Stello}, {Bruntt}, {Kjeldsen},
  {Bedding}, {Arentoft}, {Gilliland}, {Nuspl}, {Kim}, {Kang}, {Koo}, {Lee},
  {Sterken}, {Lee}, {Jensen}, {Jacob}, {Szab{\'o}}, {Frandsen}, {Csubry},
  {Dind}, {Bouzid}, {Dall}, \& {Kiss}}]{2007MNRAS.377..584S}
{Stello} D. {et~al.}, 2007, \mnras, 377, 584

\bibitem[{{Stetson}(2000)}]{2000PASP..112..925S}
{Stetson} P.~B., 2000, \pasp, 112, 925

\bibitem[{{Taylor}(2007)}]{2007AJ....133..370T}
{Taylor} B.~J., 2007, \aj, 133, 370

\bibitem[{{van den Berg} {et~al}\mbox{.}(2002){van den Berg}, {Stassun},
  {Verbunt}, \& {Mathieu}}]{2002A&A...382..888V}
{van den Berg} M., {Stassun} K.~G., {Verbunt} F., {Mathieu} R.~D., 2002, \aap,
  382, 888

\bibitem[{{VandenBerg} \& {Stetson}(2004)}]{2004PASP..116..997V}
{VandenBerg} D.~A., {Stetson} P.~B., 2004, \pasp, 116, 997

\bibitem[{{Yadav} {et~al}\mbox{.}(2008){Yadav}, {Bedin}, {Piotto}, {Anderson},
  {Cassisi}, {Villanova}, {Platais}, {Pasquini}, {Momany}, \&
  {Sagar}}]{2008A&A...484..609Y}
{Yadav} R.~K.~S. {et~al.}, 2008, \aap, 484, 609

\bibitem[{{Yakut} {et~al}\mbox{.}(2009){Yakut}, {Zima}, {Kalomeni}, {van
  Winckel}, {Waelkens}, {De Cat}, {Bauwens}, {Vu{\v c}kovi{\'c}}, {Saesen}, {Le
  Guillou}, {Parmaks{\i}zo{\u g}lu}, {Ulu{\c c}}, {Khamitov}, {Raskin}, \&
  {Aerts}}]{2009A&A...503..165Y}
{Yakut} K. {et~al.}, 2009, \aap, 503, 165

\bibitem[{{Zechmeister} \& {K{\"u}rster}(2009)}]{2009A&A...496..577Z}
{Zechmeister} M., {K{\"u}rster} M., 2009, \aap, 496, 577

\bibitem[{{Zhao} {et~al}\mbox{.}(1993){Zhao}, {Tian}, {Pan}, {He}, \&
  {Shi}}]{1993A&AS..100..243Z}
{Zhao} J.~L., {Tian} K.~P., {Pan} R.~S., {He} Y.~P., {Shi} H.~M., 1993, \aaps,
  100, 243

\end{thebibliography}

\bsp	
\label{lastpage}
\end{document}